# A brief history of the solar diameter measurements: a critical quality assessment of the existing data.

Jean Pierre Rozelot[(1)] Alexander G. Kosovichev[(2)] & Ali Kilcik[(3)]

[(1)] Université de la Côte d'Azur (CNRS-OCA, Nice) & 77 ch. des Basses Moulières, 06130 Grasse (F)
E-mail: jp.rozelot@orange.fr

[(2)] New Jersey Institute of Technology, Newark, NJ 07103, USA
E-mail: sasha@bbso.njit.edu

[(3)] Akdeniz University, Faculty of Science, Dpt of Space Science & Technologies, 07058, Antalya (T)
E-mail: alikilcik@akdeniz.edu.tr

**Abstract**

The size of the diameter of the Sun has been debated for a very long time. First tackled by the Greek astronomers from a geometric point of view, an estimate, although incorrect, has been determined, not truly called into question for several centuries. The French school of astronomy, under the impetus of Mouton and Picard in the XVIIth century can be considered as a pioneer in this issue. It was followed by the German school at the end of the XIXth century whose works led to a canonical value established at 959''.63 (second of arc). A number of ground-based observations has been made in the second half of the XIXth century leading to controversial results mainly due to the difficulty to disentangle between the solar and atmospheric effects. Dedicated space measurements yield to a very faint dependence of the solar diameter with time. New studies over the entire radiation spectrum lead to a clear relationship between the solar diameter and the wavelength, reflecting the height at which the lines are formed. Thus the absolute value of the solar diameter, which is a reference for many astrophysical applications, must be stated according to the wavelength. Furthermore, notable features of the Near Sub-Surface Layer (NSSL), called the leptocline, can be established in relation to the solar limb variations, mainly through the shape asphericities coefficients. The exact relationship has not been established yet, but recent studies encourage further in-depth investigations of the solar subsurface dynamics, both observationally and by numerical MHD simulations.

1. **Following Greek and Islamic scholars**

From time immemorial men have striven to measure the sizes of celestial bodies and among them the solar diameter holds an important place. Aristarchus of Samos (circa 310-230 BC), by a brilliant geometric procedure was able to set up the solar diameter $D_\odot$ as the 720th part of the zodiacal circle, or 1800 seconds of arc ('') (i.e. 360°/720). A few years later, Archimedes (circa 287-212 BC) wrote in the *Sand-reckoner* that the apparent diameter of the Sun appeared to lie between the 164th and the 200th part of the right angle, and so, the solar diameter $D_\odot$ could be estimated between 1620'' and 1976'' (or 27'00'' and 32'56'' (Lejeune, 1947; Shapiro, 1975). These values, albeit a bit erroneous are not too far from the most recent determinations, indicating the great skillfulness of the Greek astronomers. Curiously, such values were not truly questioned during several centuries, except during the fourteen century under the impetus of the Marãgha School (Iran) when Ibn-al-Shatir (1304-1375) wrote in his book untitled *"The Final Quest Concerning the Rectification of Principles"* that the solar diameter varies from 29'.5'' (at apogee) to 32'.32'' (at perigee), a fact that obviously Greek astronomers could not evoke in their geometric demonstration. The ratio found by Ibn-al-Shatir, of 0.913, is still erroneous, as the best present estimate is 0.967, but not too much (6 %). A bit later on, in 1656, Giovanni Battista Riccioli (1598-1671) in reviewing the measures of the solar diameter, reported a lower limit of 30'.30'' (given by Kepler) and an upper limit of 32'.44'' (given by Copernicus). Lastly, the French school led by Gabriel Mouton (1618-1694) and Jean Picard (1620-1682) can be considered as a pioneer in this field, as the two astronomers were able to report the first solar radius measurements with a modern astrometric accuracy (0.8 %)[1].

Major results of the solar semi-diameter as obtained from telescope observations from 1667 to 1955 have been compiled by Wittmann (1977). A more complete history of the solar diameter determinations, at least up to the year 2011, can be found, for example, in Rozelot & Damiani (2012).

2. **The solar diameter: a not so obvious parameter**

The solar sphere is not a stainless steel ball, for which the diameter would be easy to define. As the Sun is a fluid body in rotation, the passage of the flow from the interior to the gaseous external layers is progressive and the solar radius needs a definition. The most commonly used definition stipulates that the radius is defined as the half distance between the inflection points of the darkening edge limb function, at two opposite ends of a line passing through the disk center. Other definitions could be used, for instance the location of the limb at the minimum of temperature, or an equipotential level of gravity which defines the outer shape, etc.

From the ground, measurements of the solar radius suffer from different flaws. One of the main effect is due to the terrestrial atmospheric disturbances. When the solar ray travels from the upper to lower layers of our atmosphere, it can be affected by various factors such as moving pockets of air masses, changes in the refractive index, winds or jet streams (Badache-Damiani et al., 2007), etc. The other effects are scintillation and blurring (Rösch & Yerle, 1983). The convolution of the edge of a uniformly bright half-plane by a spread function having a center of symmetry gives a smooth symmetrical profile. In such a case, the inflection point is exactly on the edge of the object, whereas if the object is limb-darkened, this inflection point is shifted inside increasingly with the width of the spread function. Except rare cases, such effects have been, up to now, poorly taken into account, leading to spurious ground-based measurements. The past (ground-based) radius data can be used for historical studies, but must be carefully examined for astrophysical purposes. Even in the case of space measurements, major problems may occur, such as thermal or misalignment effects of the instruments in space. Thus, there is still a need for more accurate measurements.

3. **On the necessity to measure the solar radius $R_\odot$ with a high accuracy**

**3.1. In astrophysics (and geophysics),** the diameter of the Sun is a fundamental parameter that is used in physical models of stars. First of all, the diameter of stars are defined relative to that of the Sun. Therefore, a change in the absolute value of the solar diameter, as well as its temporal variations, if any, could have an impact on the inferred stellar structures. Secondly, concerning our star, a diameter estimate permits to compute the amount of energy transmitted to the Earth. The total radiative output of the Sun establishes the Earth's radiation environment and influences its temperature and atmospheric composition. Recent studies indicate that small but persistent variations in the solar energy flux may play a significant role in climate changes through direct influence on the upper terrestrial atmosphere (see for instance Suhhodolov et al., 2016).

---

[1] As an example, Picard found $R_\odot$ = 964''.8 on March 9, 1670, a remarkably accurate value.

**3.2. In solar physics,** a change in the solar size is indicative of a change in the potential energy which could be driven by such means. To the first order, a change in the radius of the Sun causes changes of its luminosity, according to the Stephan's law: $L = (4\pi R_\odot^2)\, \sigma T^4$, which gives $\Delta L/L = 4\, \Delta T/T + 2\, \Delta R_\odot/R_\odot$, where $L$ is the solar irradiance, and $T$ the solar effective temperature. Taking $L$ = 1361 W/m$^2$, $T$ = 5772 K (as recommended by the IAU) and $\Delta L/L$ = 0.01 %, it turns out that $\Delta R_\odot$ = 9.4 mas (6.8 km) if $\Delta T$ = 1.35 K over the solar cycle as found by Caccin & Penza (2003). Note that if the quiet Sun were immutable as suggested by Livingston (2005), $\Delta T \approx 0.$ K, there would be neither sunspots nor faculae, hence $\Delta L/L \approx 0.$ % as the irradiance variability reflects their presence (see for example Krivova et al., 2003). In such a case, $\Delta R_\odot$ would be $\approx 0.$ mas.

This result ($\Delta R_\odot$ in any case is less than 15 mas) is not surprising. Dziembowski et al. (2001) calculated a photospheric radius shrinkage of about 2-3 km/year with the rising solar activity. Goode et al. (2003) using a helioseismology analysis of high-degree oscillation modes from SOHO/MDI (Scherrer et al., 1995) found a shrinking of the solar surface/convection layer (which seems to be cooler) with the increasing activity, at a level consistent of the direct radius measurements based on the SOHO/MDI intensity data. Lastly, using a self-consistent approach taking into account the solar oblateness, Fazel et al. (2008) obtained an upper limit on the amplitude of the cyclic solar radius variations (a non-homologous shrinking) between 3.87 and 5.83 km, deduced from the gravitational energy variations (see also section 7). Such results rule out all observed variations which can be found in several papers claiming that the solar diameter may change in time by some 300 mas or more.

**3.3. Precise limb shape (curvature) changes** both in latitude and time, due to an aspherical thermal structure. Such alterations play a role in the physics of the solar sub-surface layers. According to the observed temporal variation of f-mode frequencies, the near sub-surface solar layer (NSSL) is stratified in a thin double layer, interfacing the convective zone and the surface (Lefebvre & Kosovichev, 2005). This shear layer called "*leptocline*" -from the Greek "*leptos*": thin and "*klino*": hill- (Godier and Rozelot, 2001, Lefebvre et al., 2009) is the seat of many phenomena: an oscillation phase of the seismic radius, together with a non-monotonic expansion of this radius with depth, likely an inversion in the radial gradient of the rotation velocity rate at about 50° in latitude, opacities changes, superadiabicity, the cradle of hydrogen and helium ionisation processes and probably the seat of in-situ magnetic fields (Lefebvre et al., 2006; Barekat et al., 2016).

A possible mechanism has already been stated as followed (Pap et al., 2001): although the ultimate source of the solar energy is the nuclear reactions taking place in the center of the Sun, the immediate source of the radiating energy is the solar surface. The nuclear reactions are almost certainly constant on the time scales shorter than millions of years, but the mechanisms which carry the energy to the solar surface may not be. Indeed, observations of the solar radiation integrated over the entire solar spectrum (total irradiance), obtained by space-based experiments now over several decades, have demonstrated that the total irradiance varies on the time scales from minutes to the 11-yr solar cycle. If the central energy source remains constant while the rate of energy emission from the surface varies, there must be an intermediate reservoir where the energy can be stored or released depending on the variable rate of energy transport. The gravitational field of the Sun is one such energy reservoir. If the energy is stored in this energy reservoir, it will result in a change in the solar radius. Thus, a careful determination of the time dependence of the solar radius can provide a constraint on models of total irradiance variations.

Recent analysis of the high-degree oscillation modes revealed a sharp gradient of the sound speed in a narrow 30-Mm deep layer just beneath the solar surface (Reiter et al., 2015). The complex physics of this near surface shear layer (the leptocline) presumably plays an important role in the solar dynamo (Pipin & Kosovichev, 2011). To this respect, new features of the Solar Dynamics Observatory/Helioseismic and Magnetic Imager (SDO/HMI) analysis is that the HMI data allow us to reconstruct the flows in this shallow subsurface layer, and match these to the directly observed surface flows (Fig. 1). Such flows maps permit to investigate other important properties of the subsurface dynamics of the Sun, which previously were not accessible (e.g. Kosovichev & Zhao, 2016).

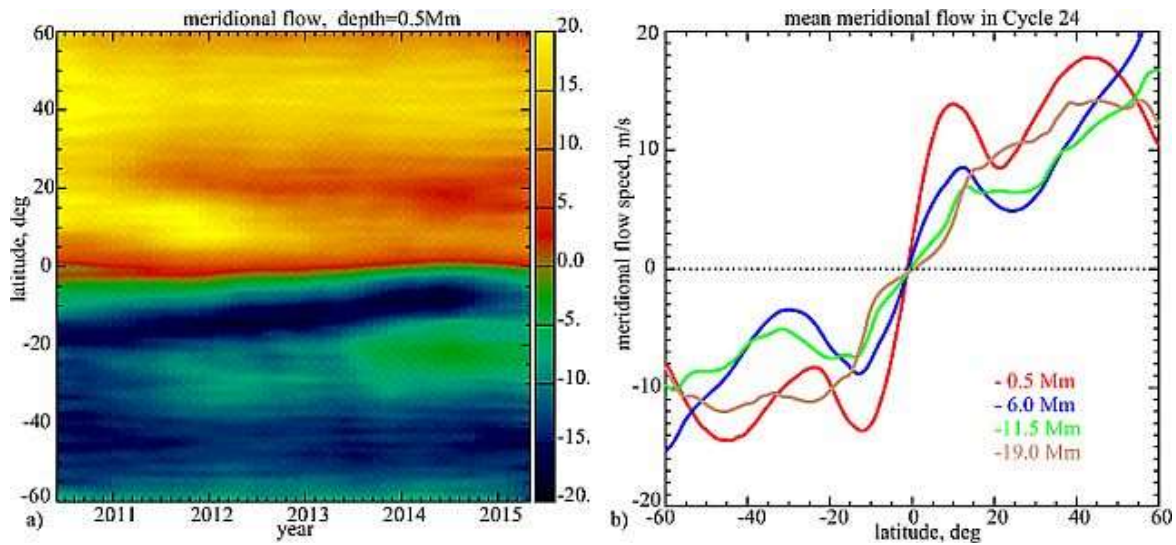

**Figure 1.** Left (a): Evolution of the subsurface meridional flows obtained from the 5-years of the SDO/HMI observations during Solar Cycle 24. The red and yellow colors show the flow components towards the North pole, the green and blue colors show the South-ward flow. The color scale range is from -20 to 20 m/s. Right (b): The mean meridional flow averaged for the whole period of observations at four different depths, showing the importance of the Near Sub-Surface Layers (NSSL, or *leptocline*). After Kosovichev & Zhao (2016).

**3.4. Temporal solar size variations,** even faint, imply changes of the gravitational moments, $J_2$ and $J_4$. Let us recall that the gravitational moments are linked to both the distribution of mass and the (differential) rotation of the body, from the core to the surface. Precise knowledge of such gravitational moments is required to develop high precision astrometry and in addition, may constraint gravitational theories, both from a theoretical and experimental points of view. In such prospect, the Eddington-Robertson parameters, $\gamma$, and $\beta$ contributes to the relativistic precession of planets. Note that $\gamma$ encodes the amount of curvature of space-time per unit rest-mass, and the post-Newtonian parameter $\beta$ encodes the amount of non-linearity in the superposition law of gravitation. In the case of the solar system, it is still difficult to disentangle $J_2$, $\gamma$ and $\beta$. However, by accurately measuring the limb curvature over the latitudes -that is to say the solar shape (and to first order the oblateness)-, it is possible to get a good estimate of the solar quadrupole moment, to an accuracy of one part in 200 of its size of around $10^{-7}$. Recent analysis includes the Lense-Thirring precession effect, which is not negligible. In the case of Mercury for instance, it may have been canceled to a certain extent by the competing precession caused by a small mismodeling in the quadrupole mass moment $J_2$ of the Sun (Iorio, 2011).

### 4. A quick tour of solar diameter measurements

*4.1. The so-called Danjon Astrolabe,* was redesigned in the early 1970's in a solar astrolabe in order to get measurements of the diameter of the Sun[2]. After a protecting glass density at the front window of the instrument, the image is split by a reflector prism and a mercury mirror and then focused by a refracting telescope. In the case of a perfect setting, the direct and reflected images are symmetric in relation to the axes of a reticule. The time of their coincidence in the center of the eyepiece field corresponds to the exact moment when the observed Sun edge crosses the parallel of altitude corresponding to the angle of the prism. In principle, this kind of apparatus is suited for solar observations and series of measurements were obtained, first in France showing a periodicity in the radius data of about 900 days, in antiphase with the solar activity.

This question of the time dependence of the solar radius had already been debated by Secchi and Rosa (1872)[3] and numerous other observers. However, it seems that Cimino (1944) following previous works made by Gialanella (1941)[4] was the first to put in evidence two oscillations in the solar observations made in Roma (I) from 1876 to 1906: a first one of 22.5-yrs, of amplitude 0''.2 - 0''.5, for a mean radius of 961''.30, attributed to the solar variability, and a second one of (6.5-8.5) -yrs, that the author did not identified with a solar mechanism, ascribing it rather to an atmospheric effect.

Giannuzi (1953, 1955), in her analysis of the solar radius observed at Greenwich (UK) from 1851 to 1937, confirmed the basic cycle of 22 to 23-yrs, modulated by a shorter one of about 7.5 -yrs, in anti-phase to solar activity, "*leading* -according to the author- *to a great suspicion of its atmospheric origin*". This last periodicity is three times longer than the one found in the Calern data i.e. ≈ 900 days (Laclare 1983; Débarbat & Laclare (1999), as 7.5 yrs * 365.25 is 2739 days; dividing by 900 this gives ≈ 3.0(4), so that such estimates are harmonics. In any event, disregarding the Gianuzzi publications, the CERGA's results inflamed the scientific community. The solar astrolabe was duplicated and used in Algeria, Brazil (Rio de Janeiro and São Paulo), Chile, Spain and Turkey. However, the results were disparate and it was not possible, in spite of considerable efforts, to conciliate the data obtained in different places. Moreover, the amplitude of the modulation found for the cyclic variation of the solar radius with time was so high[5] that no theory was able to explain them. On a pure statistical basis, without knowing the part due to the atmospheric fluctuations and the part due to the solar component, it was found from the photoelectric astrolabe (DORAYSOL, installed at Calern -F) a mean solar radius of (959.48 ± 0.32)'' as deduced from 19169 measurements between the years 1999 and 2006, two times more measurements than with the visual astrolabe during 26 years, from 1978 to 2004. By comparison, over the same period (1999-2006), a mean solar radius of (959.55 ± 0.01)'' was deduced from 371 solar visual astrolabe observations (Morand et al. 2010).

The Picard-sol program implemented also at Calern (F), led to a solar radius of (959.78 ± 0.19)'' (or 696 113 ± 138 km) at 535.7 nm (Meftah et al., 2014) between 2011 and 2013. Such inconsistent results which are supposed to measure the same astronomical object, obtained at the same site, during similar periods of time, show evidently a mix of atmospheric and solar signals, and may reinforce what was earlier detected in Roma. Lastly, temporal variations over the solar cycle, larger than 0''.5, found in the astrolabe data are incompatible with modern relativistic theories. The network of the solar astrolabes was progressively abandoned.

---

[2] The principle of the astrolabes have been described in Kovalevsky J., Lecture Notes in Physics: Astrométrie Moderne, 1990, Vol. 358, Springer (Heidelberg-D), chapter 7, pp 173-194.
[3] The so-called Sechi-Rosa law stipulates that the solar radius variability is out of phase of the sunspot activity.
[4] He reported a mean radius of 961''.38 from observations made (by means of a meridian telescope) at Campidoglio and Monte Mario (near Roma, I), from 1876 to 1937 pointing out a possible effect of the atmosphere. See *Commentationes, Accademico Ponteficio Giuseppe Armellini nella Tornata*, November, 30, Vol. VI, 25, pp. 1142-1200.
[5] Some 0''.7 at the beginning of the observations, a value which was progressively reduced down to 0''.5 twenty years later.

*4.2. Several other radius measurements were made from the grounds which cannot be all reported here. To be mentioned:*

- **The drift-time measurements** of the solar diameter made with two optically identical 45-cm aperture solar telescopes at Izaña (SP) and Locarno (CH) by Wittmann & Bianda (2000, and references herein). Their last results show that the radius is (960.63 ± 0.02)'', from 7583 visual transit observations made in Izaña during 1990-2000 and (960.66 ± 0.03)'', from 2470 visual transits made in Locarno during 1990-1998, both at the wavelength about 550 nm. The data show no long term variations in excess of about 0.0003''/yr and no cycle-dependent variations in excess of about ± 0''.05.

- **The Mt Wilson (USA) system** (by spectrography) measures the Sun's apparent radius in the neutral iron spectral line at 525 nm (Ulrich & Bertello, 1995). The radius measured by the authors is derived from the drop-off of brightness of the solar disk at its edge and is a function of the temperature and density profile. This definition is different from the usual one based on the inflection points and, moreover, the measurements are made using a spectral line and not in white light. From 1982 to 1994, the authors found a time dependence of the solar radius in phase with the solar activity, in contradiction with the CERGA's observations in France. Analysis of the Mt Wilson (USA) data over 30-yrs were completed by Lefebvre et al. (2006) leading to the conclusion that there is no clear correlation between the temporal variations of the apparent solar radius and the variations in the sunspot number. However, the data show a distortion of the observed apparent solar figure: a bulge appears near the equator extending through 20°-30° of heliographic latitudes, followed by a depression at higher latitudes. The global behavior of the shape remains oblate. This result is qualitatively consistent with other measurements taken at the Pic du Midi (F) observatory. From a physical point of view, this distorted shape of the Sun reflects the influence of the gravitational moments ($J_2$ and $J_4$).

- **The Solar Diameter Monitor (SDM)**, a dedicated instrument} at the High Altitude Observatory (USA), to measure the duration of solar meridian transits during the 6 years 1981-1987, spanning the declining half of solar cycle 21. The Brown's (1987) report stated that for this period, the annual averages of the diameters differed from each other by less than 0''.05. Such results seem to be contradictory to the Bertello and Ulrich measurements, albeit one set is obtained using a single spectral line and the other one in white light. Combining the SDM photoelectric measurements with models of the solar limb-darkening function, Brown & Christensen-Dalsgaard (1988) were able to derive a mean value for the solar near-equatorial radius of (695.508 ± 0.026) Mm. Annual averages of the radius were found identical within the measurement error of ± 0.037 Mm.

- **The Reflecting Heliometer**}, an improved version of the solar astrolabe operating by drift scans, is operating at the "Observatorio Nacional" in Rio de Janeiro (BR) since 2011. It measures the solar diameter at all heliographic latitudes by rotating around its axis. A linear fit of the data recorded up to 2015 (about 11500 measurements per calendar year) leads to a value of the solar radius of 958''.7, with a dispersion of around ± 0''.5, more or less in phase with the solar activity (Boscardin et al. 2016). However, after a pure statistical study at a site where the seeing is rather high, no deconvolution of the atmospheric parameters has been made.

- An interesting study has been made by Hiremath (2015) in compiling the **Solar White-light Images of the Sun obtained at the Kodaikanal Observatory** (IN). Photographic data extend back to 1904 and cover 106 years. All the plates are digitized and available (31800 plates covering about 31000 days). The solar radius has been extracted after removing the limb darkening using an original method. Results, still in progress, seems to show that during the years 1923-1945, the Sun's radius is constant and does not change with the solar cycle.

- **Historical archives of the Royal Observatory of the Spanish Navy** (today the "*Real Instituto y Observatorio de la Armada -ROA-*" located at Cadiz -SP) has been recently reanalyzed by Vaquero et al. (2016) in the scope to recover the solar radius measurements since 1753, i.e. during the past 250 years. From this long-term perspective, the data do not present any significant trend from the statistical point of view, and if any, it would be within the measurement uncertainty. Thus, the Spanish solar observations show that the solar diameter did not change in the past 250 years. The mean value has been estimated after applying corrections for refraction and diffraction to (958.87 ± 1.77) ''.

**4.3. The first measurements in the near outer atmosphere** were made by means of a dedicated instrument embarked in a balloon nacelle (Sofia et al. 1991). The primary goal was to determine the solar oblateness, with a precision of $\approx 10^{-5}$, for which it was envisaged at that time to test the Brans-Dicke theory (see a review in Damiani et al., 2011).

Seven flights were made in the Arizona desert (USA) spanning the years 1992 to 2011 (first flight in May 4, 1990 at an altitude of 120 000 feet). Data have been several times revisited. The last values are listed in Table 1 (Sofia et al., 2013). Inspection of this table shows how it is difficult to measure the solar radius, as the amplitude variation with time is 244 mas, about 10 times larger than the one deduced from satellite observations (see section 6). Moreover, the records permitted to deduce the solar oblateness (last column in Table 1), but the SDS-1992 estimate is not in agreement with inferences based on the rotation at the solar surface, and far below (of expected $7.8*10^{-6}$ for a uniform rotation; and the differential rotation at the surface increases the oblateness (see a discussion in Rozelot et al., 2009).

| Flight number | Year | $R_\odot$ at 1 AU in '' | Oblateness $\Delta r$ in mas |
|---|---|---|---|
| 6 | 1992.82 | 959.638 ± 0.020 | 4.13 ± 1.92 |
| 7 | 1994.81 | 959.675 ± 0.020 | 8.16 ± 2.02 |
| 8 | 1995.82 | 959.681 ± 0.020 | 8.25 ± 1.34 |
| 9 | 1996.85 | 959.818 ± 0.020 | 9.88 ± 1.82 |
| 10 | 2001.83 | 959.882 ± 0.040 | |
| 11 | 2009.87 | 959.750 ± 0.020 | |
| 12 | 2011.86 | 959.856 ± 0.020 | |

Table 1. Summary of the Solar Radius Observations made by means of the balloon SDS experiment given here as an example, the amplitude being 244 mas, about 10 times larger than the one deduced from satellite observations. The last column displays the solar oblateness, i.e. the difference between the equatorial and polar radius in millisecond of arc (mas).

## 5. Solar Diameter from Eclipse data

The diameter of the Sun can be measured during eclipse times, a technique that covers now about three centuries. Such observations present, in theory some advantages since atmospheric effects are less important than in any other ground-based solar observation techniques. The main difficulty lies in the complexity of the lunar profile (due to the presence of valleys and craters) superimposed on the solar limb, and the technique depends upon the libration of the Moon. Another difficulty arises from the limited number of data obtained because of the few events per year. However, because of the short eclipse duration and the fast-moving trajectory in space, total and annular solar eclipse observations can provide more accurate calibration points by comparison to other data (Fiala et al., 1994; Dunham et al., 2005; Lamy et al., 2014).

The oldest reliable observation goes back to 1715 (Danylevsky, 1999), and the interest in solar eclipses for estimating the solar diameter increased in the early 1980s (Dunham et al., 1980). Several tables listing the solar radius measurements from solar eclipses since 1715 are available, such as in Dunham et al. (2016) or in Kilcik et al. (2009). As an example, the last authors determined the solar radius of (959.22 ± 0.04)'', calibrated to 1 AU. Lamy et al. (2014), averaging a set of 17 estimations at the 2010, 2012, 2013 and 2015 eclipses found a solar radius of (959.99 ± 0.06)'' (or 696 246 ± 45 km according to the authors). New measurements are scheduled for the next total solar eclipse in 2017 (August 21), for which the path of totality is very favorable for observations.

## 6. Solar diameter from satellite missions

Historical data suggest that the Sun's radius may have changed over short or long periods of times. But, as seen in the previous sections, the question remained open at the eve of satellite missions. The ground-based contradictory results, as well as the difficulty to measure the solar radius from balloons, highlighted the necessity of more sensitive space experiments. The first mission of that kind was provided by the Michelson Doppler Imager (MDI) instrument (Scherrer et al. 1995), on board the Solar and Heliospheric Observatory (**SOHO**) satellite[6]. It offered for the first time very accurate solar radius measurements from space. The MDI results had been analyzed by Emilio et al. (2000, 2004), who found a variation within the solar cycle of $dr_{cycle}$ = (+ 21 ± 3) mas, 5 times smaller than the best ground-based measurements, and furthermore, in phase with the solar activity. A re-analyis of the data, taking into account instrumental corrections (Bush et al., 2010) led to the conclusion that any intrinsic changes in the solar radius, *that are synchronous with the sunspot cycle*, must be smaller than 23 mas, peak-to-peak. In addition, the authors find that the average solar radius must not be changing (on average) by more than 1.2 mas yr$^{-1}$. According to Kuhn (2004), if ground -and space- based measurements are both correct, the pervasive difference between the constancy of the solar radius seen from space and the apparent ground-based solar astrometric variability can only be accounted for by long-term changes in the terrestrial atmosphere.

---

[6] Section 2.3.10 of this paper stipulated that MDI will make the first photometric observations of the complete solar limb from above the atmosphere, determining the shape of the solar disk to an accuracy of about 0.''0007 each minute.

Adopting the value 1 AU = $1.495979 \times 10^8$ km, the MDI observations gives the Sun's radius $R_\odot = (6.9574 \pm 0.0011) \times 10^5$ km. This is slightly smaller than the Brown & Christensen-Dalsgaard (1998) measurements (see above), but is consistent with the highly precise helioseismic determination made by Schou et al. (1997) of $(6.9568 \pm 0.0003) \times 10^5$ km. The difference between what is called the "*seismic*" and the "*photospheric*" radius of $(0.347 \pm 0.006)$ Mm with respect to $\tau_{5000} = 1$, has been explained by the difference between the height at disk center (where $\tau_{5000} = 1$) and the inflection point of the intensity profile on the limb (Habeirreter et al. 2008).

Early as in 1996, a group of solar physicists led by Damé, Rozelot and Thuillier (Damé et al. 2000) proposed a dedicated mission focused on the accurate measurement of the solar diameter from space, which was accepted by the French Agency CNES as the **PICARD** mission, and launched in 2010. Unfortunately, the influence of the space environment (ultra-violet radiation, thermal cycling, etc), led to considerable degradation of the instruments in orbit (contamination, temperature variations, abnormal detector responses, etc) and the mission was shortened. However, the data were analyzed by Meftah et al. (2015) who found that changes in the solar radius amplitudes obtained by the PICARD space telescope were less than ± 20 mas (i.e. ± 14.5 km) during the years 2010-2011. Considering this short period of time, the observations could not provide any direct link between solar activity and significant fluctuations in the solar radius. The authors concluded that the variations, if they exist, are within the range of values, which could be modulated by a typical periodicity of 129.5 days with a ± 6.5 mas variation[7].

Other space missions were used in attempts to measure the solar diameter, although they were not specifically designed for such a purpose. Indeed they were more focused in determining the shape of the Sun, its oblateness at first. The distortion of the limb under the influence of the gravitational moments provides a ``true figure'' of the Sun. Such a topic, "*how round is the Sun*" could be another story.

**1.** The primary goal of the **RHESSI** (Reuven Ramaty High Energy Solar Spectroscopic Imager) mission was to explore the basic physics of particle acceleration and explosive energy release in solar flares, but it was used also to measure the solar limb using the optical solar aspect sensor. The results had been described by Fivian et al. (2008) and Hudson & Rozelot (2010), for the oblateness part, and in Battaglia & Hudson (2014) concerning the limb heights. Their first attempts to determine a solar radius in X-ray observations sound good, but a more detailed analysis remains to be done. Nevertheless, an extrapolation of the measured points shown on the left part of the curve given in Fig. 4[8] leads to $R_\odot \approx 969''.0$, at 1 nm. It would be useful to explore this issue in future solar limb measurements by RHESSI.

**2**. The **Solar Dynamics Observatory** (SDO) satellite was launched on February 11, 2010, with a specific program designed to understand the causes of solar variability and its impacts on Earth. Among a broad spectrum of scientific themes, two were specifically mentioned at the beginning of the mission: the study of the impact of active regions on the solar diameter and the monitoring of a deep Sun survey (using an imaging spectropolarimeter, Helioseismic and Magnetic Imager -HMI-, Scherrer al., 2012) at 617.3 nm dedicated to helioseismology and magnetic field study). With regard to the measurements of the solar diameter, results have been analyzed by Emilio et al. (2015) during the Venus transit of 2012[9]. They found in the continuum wing of the 617.3 nm line by means of the HMI instrument, a solar radius at 1 AU of (959.57 ± 0.02)'' (or 695 946 ± 15 km). The AIA instrument observed simultaneously the Venus transit at ultraviolet wavelengths which gave (963.04 ± 0.03)'' at 160.0 nm and (961.76 ±0.03)'' at 170.0 nm. Oblateness measurements have been substantially analyzed by Meftah et al. (2016) who employed several descriptions and analysis already made in their previous publications.

---

[7] Shall we consider this period as a sub-harmonic of the 7.5-8.5 years periodicity found by Cimino as previously quoted (i.e.: $8.5 \times 365.25/129.5 \approx 24$) ?

[8] See Rozelot et al., 2015, Fig. 3: $R_\odot = 4.426 \, 10^{-5} x^2 - 4.226 \, 10^{-2} x + 969.056$, where *x* is the wavelength in nm.

## 7. Helioseismic measurements

The significance of the actual number of the solar radius (and we need to know it precisely- see section 3) lies to a large extend on our increasing knowledge in helioseismology (see for example Ehgamberdiev, S. this book, the Standford site: http://soi.stanford.edu/results/heliowhat.html, or Kosovichev, 2011 for a comprehensive review).

It was only in 1997 that Schou et al. (1997) succeeded for the first time, in obtaining a helioseismic determination of the solar radius by using high-precision measurements of oscillation frequencies of the f-modes of the Sun, obtained from the MDI experiment on board the SOHO spacecraft. They determined a "seismic radius" of about 300 km smaller than the photospheric adopted radius. A similar conclusion was reached by Antia (1998) on the basis of analysis of data from the GONG network. The question remained open until an attempt to reconcile the two values by Haberreiter et al. (2008) as previously seen.

As pointed out by Di Mauro (2003), the helioseismic investigation of the solar radius is based on the principle that the frequencies of the f modes of intermediate angular degree depend primarily on the gravity and on the variation of density in the region below the surface, where the modes propagate. It can be shown from the asymptotic dispersion relation[10] that $\omega \propto R_\odot^{-3/2}$ and by applying a variational principle, one can deduced a simple relation between f-mode frequencies, $\omega_{l,0} = 2\pi\nu_{l,0}$ , so that the correction $\Delta R$ that has to be imposed to the photospheric radius $R_\odot$ assumed for the standard solar model is:

$$\frac{\Delta R}{R_\odot} = -\frac{2}{3}\left\langle \frac{\Delta\nu_{l,0}}{\nu_{l,0}} \right\rangle ,$$

where $\langle \ \rangle$ denotes the average weighted by the inverse square of the measurement errors.

---

[9] Let us recall the observations made by the French astronomer Le Gentil in his several attempts to measure the solar diameter through the Venus transits. He reported 979.''55 during the passage of Venus on December 21rst, 1768 and 946''.75 on June 21rst, 1769, that is a mean of 963''.15, an estimate not too far from the real value. In Le Gentil, G.J. 1779, "*Voyage dans les mers de l'Inde fait par ordre du Roi*", Vol. I, Paris, Imprimerie Royale, p. 505 and 509.

[10] The dispersion relation of the f-modes is: $\omega^2 \simeq g_\odot k_h$ where $k_h = [l(l+1)]^{1/2}/R_\odot$ is the horizontal component of the wave number and $g_\odot = GM/R_\odot^2$.

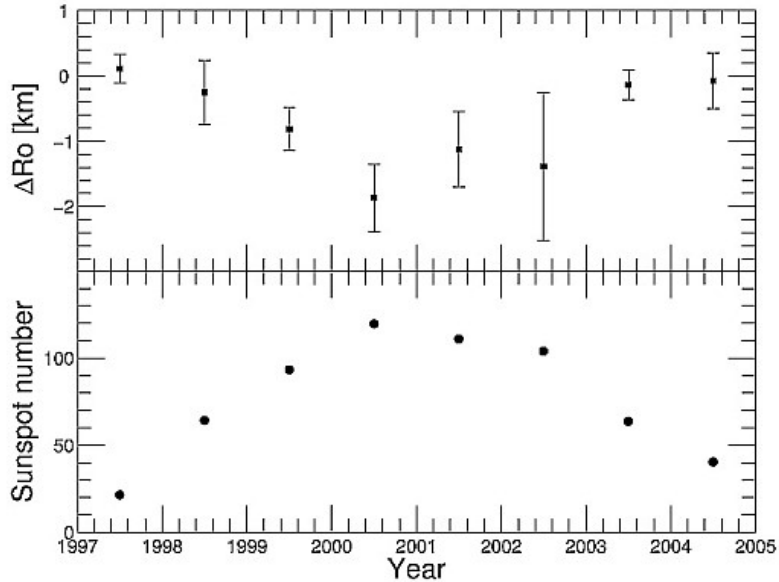

**Figure 2.** Top, temporal variation of ΔR near the solar surface at $r$ = 1 $R_\odot$; bottom, variation of the sunspot number for the same period. The variation of the seismic radius at the surface is found to be in antiphase with the solar cycle, with an amplitude of about 2 km. Computations have been made for $l$ ranging from angular degree 125 to 285 with the reference year 1996; without information at higher $l$ the surface radius cannot be constrained better. After Lefebvre & Kosovichev (2005) and Lefebvre et al. (2007).

Dziembowski et al. (1998), inferring a relation between the f-mode frequency and radius variations in the subsurface layers, with the aim of determining a possible solar cycle dependence, analyzed first the period from May 1996 to April 1997. They show that the maximal relative variation of the solar radius during the observed period was about $\Delta R/R_\odot = 6 \times 10^{-6}$, which corresponds to approximately $\Delta R$ = 4 km. In a second step, Dziembowski et al. (2000) analyzed a larger set of data spanning a period from mid-1996 to mid-1999, and they found that the systematic trend of $\Delta R/R_\odot$ was not correlated with the magnetic activity.

However, such results have been obtained assuming that $dr/r$ is constant with depth. Lefebvre & Kosovichev (2005) and Lefebvre et al. (2007) re-analyzed the time series 1996-2005 and showed that helioseismic radius varies in anti-phase with solar activity (Fig. 2) in the outer region of the Sun, but involving a change in behavior in deeper layers, the radius being non-homologous in the subsurface layers[11]. Such radius variations can give a real insight into changes of the Sun's subsurface stratification (Lefebvre et al. 2009). From another point of view, using a self-consistent approach, assuming either homologous ($n$ = 1) or non-homologous variations ($n$ = 2, ...), Fazel et al. (2004, 2009) calculated $\Delta R/R$ and $\Delta L/L$ associated with the energies responsible for the expansion of the upper layer of the convection zone. They obtained an upper limit on the amplitude of cyclic solar radius variations (anticorrelated with the solar activity) between 5.8 ($n$ = 1) and 3.9 ($n$ = 2) km.

Lastly, a recent analysis of f-modes by Kosovichev & Rozelot (2016) covering the degree range $l$ up to 1200 shows an anticorrelated variation of the seismic radius with the magnetic activity (Fig. 2). In this range (*l* between 600 and 1200), the f-mode kinetic energy is concentrated within a layer of approximately 2 Mm of the solar photosphere.

---

[11] An attempt to derive a better approximation for the kernel linking the relative frequency changes and the solar radius variation in the subsurface layers has been made also by Chatterjee & Antia with no substantial conclusions: see arXiv:0810.4213v1 [astro-ph] 23 Oct 2008.

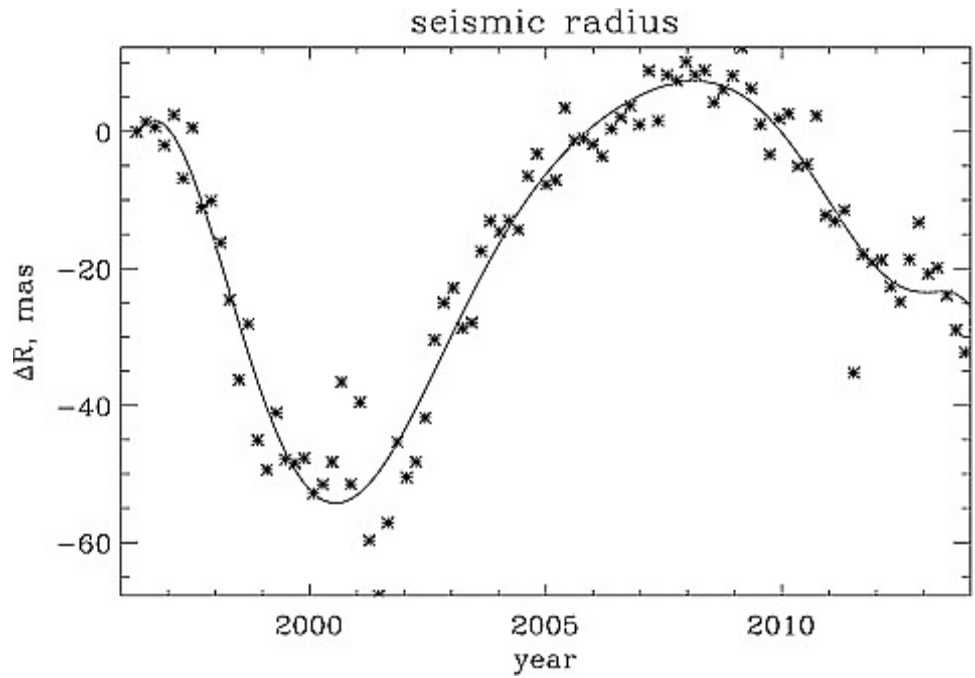

**Figure 3.** Seismic radius variation relative to the solar minimum value of 1996 as deduced from the analysis of f-modes during the years 1996-2015. The radius is clearly in opposite phase with the cycle sunspot activity. The peak-to-peak ΔR amplitude is ≈ 60 mas.

8. **Solar radius dependence with wavelength**

The importance of limb shape dependence on the wavelength, from 303 nm up to 2400 nm, was recognized since a long time (Pierce & Slaughter (1977a); Pierce et al. (1977b); Neckel & Labs (1987, 1994)). It was found that the limb-darkening function could be fitted by a fifth order polynomial with no significant variations during the solar cycle. Since then, few measurements of the solar radius variations with the wavelength have been made.

An investigation of the existing literature shows that the solar radius has been observed:

- in the UV part of the spectrum, using the Extreme Ultraviolet Imager (EIT) aboard the SOHO spacecraft and analyzed by Selhorst, Silva & Costa (2004) on the one hand, and using the AIA (Atmospheric Imaging Assembly) (Lemen et al. (2012)) instrument aboard the SDO (Solar Dynamics Observatory) satellite, during the 2012 Venus transit on the other hand;

- in the visible light spectral lines, as deduced from Mercury transits in 2003 and 2006 and from the Venus transit in June 2012, through the Michelson Doppler Imager (MDI) aboard the Solar and Heliospheric Observatory (SOHO) and from the HMI/SDO images (Helioseismic and Magnetic Imager instrument -HMI- aboard the Solar Dynamics Observatory). Sigimondi et al. (2015) observed also the Venus transit in 2004 in Athens to measure the solar radius in Hα;

- in the visible broad band continuum by the Picard-sol instrument installed at the Calern observatory, South of France (Meftah et al. (2014)[12]), and by the ``Heliometer'' instrument installed at the Pic du Midi Observatory (South of France) (Rozelot et al. (2013));

- in the radio band, several determinations of the solar radius have been made by radio telescopes at millimeter waves, including eclipse observations at centimeter and decimeter waves, and interferometric observations at meter waves (Selhorst et al. (2004); Giménez de Castro et al. (2007), (2009)).

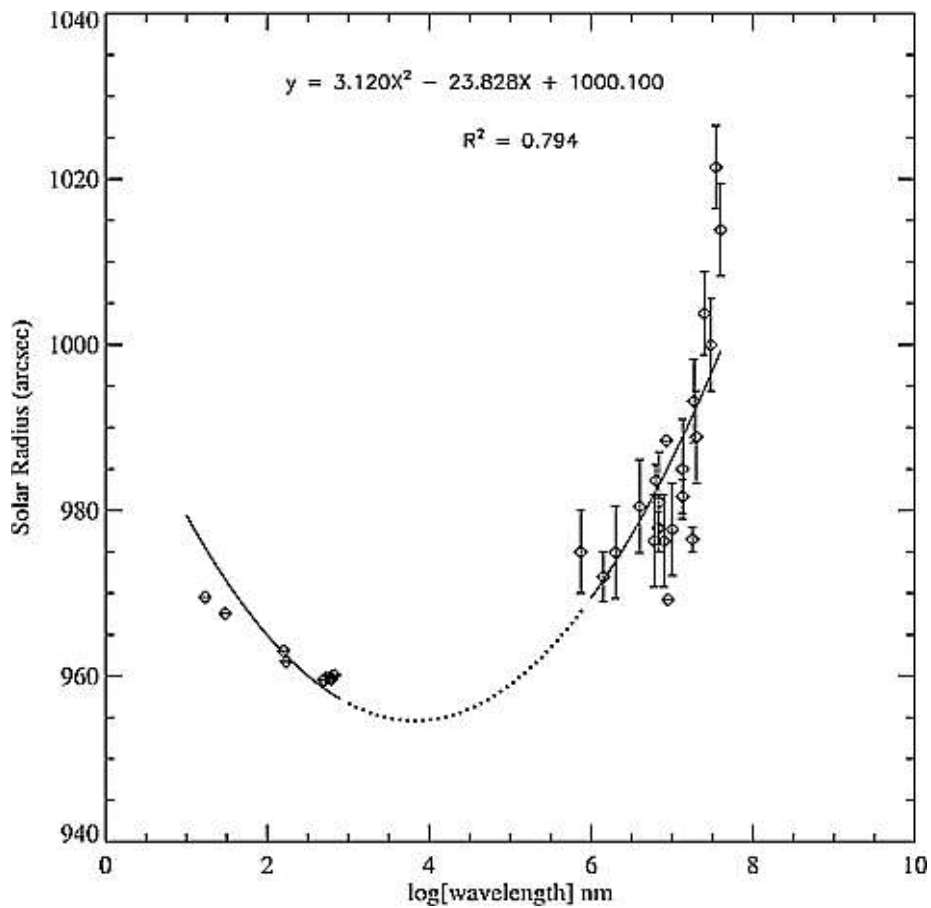

**Figure 4**. Solar radius variations from EUV to Hα (on the left side) and to millimeter radio waves (on the right side) as a function of wavelength in the decimal logarithm scale. A second order polynomial correctly fits the data showing a strong wavelength dependence of the solar radius. The mid-domain (curve in dots) ranging from 677 nm to 742,060 nm (404 GHz) is presently still unexplored. A minimum is obtained for about 6.6 μm with an estimated error of ± 1.9 μm. No unique model can currently explain such an important wavelength variation. See also Rozelot et al. (2015, 2016) for further descriptions.

Fig. 4 taken in Rozelot & Kosovichev (2015) shows all the data plotted together. A strong wavelength dependence of the solar radius is highlighted. However, a large wavelength domain from 667 to 742,060 nm is currently unexplored. In this range the polynomial fits suggests a minimum in the mid-IR region at about (6.6 ± 1.9) μm, that must be confirmed by other means (observations are scheduled with the Atacama Large Millimeter Array - ALMA- in Chile within the cycle 4 program). No model can reproduce today the entire variation from X-UV to radio. Lastly, albeit the measurements were obtained at different periods of time, no significant radius temporal variations has been found, at least at the level of the uncertainty at which the observations were made.

---

[12] A new summary of the PICARD solar radius observations at different wavelengths, based on the determination of the inflection point position, can be found in Meftah et al., 2016, SPIE Conference Paper, Vol. 9904, DOI: 10.1117/12.2232027.

## 9. Conclusion: a quest for more accurate data

This chapter shows evidently that the determination of the solar radius with a high accuracy is maybe one of the most difficult problem to solve today. The IAU General Assemblies in Pekin (2012) and in Honolulu (2015) adopted for the Astronomical Unit 1 AU = 149 597 870 700 m exactly and for the solar radius $R_\odot$ = 6.957 * $10^8$ m, so that in arc second, the radius is now fixed to 959.''22. It results a difference of 0.''41 with the previous canonical value as defined by Auwers (1891), or 297 km taking into account the new value of the Astronomical Unit. According to our graph (Fig. 4), this would lead to a nominal wavelength of 400.00 nm. In any event, further accurate observations are needed.

As far as the temporal variations of the solar radius are concerned, here also, the quest is not finished. New space dedicated satellites must be designed and launched. Such data would also help to understand the underlying mechanisms of the solar-cycle variations and physical processes in the Near Sub-Surface Layers of the Sun. Such determinations are still a challenge.

**Acknowledgment**

This work was partly supported by the *International Space Science Institute* (ISSI) in Bern (CH) where one of the author (JPR) is repeatedly invited as a visitor scientist, and also by a NASA grant NNX14AB70. JPR and AGK thanks also the *Science Development Foundation under the President of the Republic of Azerbaijan* for providing financial support to attend Baku Solar conference in June 2015.

**References**

Antia, H.M. 1998. *aap* 330, 336.

Auwers, A. 1891. *Astron. Nachr*. 128, 361.

Badache-Damiani, C., Rozelot, J. P., Coughlin, K. & Kilifarska, N. 2007. *mnras* 380, 609.

Barekat, A., Schou , J. \& Gizon, L. 2016. *aap to be published.*

Boscardin, S. C., Sigismondi, C., Penna, J. L., D'Avila, V., Reis-Neto, E. & Andrei, A.H. 2016. In ``*Solar and Stellar Flares and their Effects on Planets Proceedings*'', IAU Symposium No. 320, A.G. Kosovichev, S.L. Hawley & P. Heinzel (eds), Cambridge University Press, London (UK).

Brown, T. M. 1987. In ``*Solar Radiative Output*''. Natn. Cent. for Atmos. Res., Boulder, CO. USA, Foukal, P. ed., 176-188.

Brown, T. M., Christensen-Dalsgaard 1988. *apj* 500, L195-L198.

Bush, R.I., Emilio, M. & Kuhn, J.R. 2010. *apj* 716, 1381–-1385.

Caccin, B. & Penza, V. 2003. *Mem. S.A.It*. Vol. 74, 663-666.

Cimino, M. 1944. *Accademico Ponteficio Giuseppe Armellini, Commentationes*, Vol. VIII, 17, pp. 485-506.

Débarbat, S. & F. Laclare, F. 1990. *Acta Astronomica* 40, 313-319.

Damé L., Cugnet, D., Hersé, M., Crommelynck, D., Dewitte, S., Joukoff, A., Ruedi, I., Schmutz, W., Rozelot, J.P. et al.: 2001. In ``*Picard, solar diameter, irradiance and climate*'', Euroconference on the Solar Cycle and Terrestrial Climate, September 25-29, 2000, Tenerife, Spain, *ESA*, SP-463, 223-229.

Damiani C., Rozelot, J.P., Lefebvre S., Kilcik A. & Kosovichev, A.G. 2011. *jastp* 73, 241-250.

Danylevsky, V.O. 1999. *Contrib. Astron. Obs. Skalnate Pleso* 28, 201.

Di Mauro, M.P. 2013. In ``*The Sun's Surface and Subsurface: Investigating Shape*''. Lecture Notes in Physics, J.-P. Rozelot (ed.), Vol. 599, 31-67.

Dunham, D.W., Sofia, S., Fiala, A.D., Muller, P.M. & Herald, D. 1980. *Science* 210, 1243.

Dunham, D.W., Thompson, J.R., Herald, D.R., Buechner, R., Fiala, A.D., Warren, W.H. Jr., Bates, H.E. 2005. *SORCE Science Meeting* September 14-16, Durango, Colorado (USA).

Dunham, D.W., Sofia, S., Guhl, K. & Herald, D. 2016. In ``*Solar and Stellar Flares and their Effects on Planets*''. Proceedings IAU Symposium No. 320, A.G. Kosovichev, S.L. Hawley & P. Heinzel (eds). Cambridge University Press, London (UK).

Dziembowski, W.A., Goode, P.R. & Schou, J. 2001. *apj* 553, 897.

Dziembowski, W.A., Goode, P.R., Di Mauro, M.P., Kosovichev, A.G. & Schou, J. 1998. *apj* 509, 456.

Dziembowski, W.A., Goode, P.R., Kosovichev, A.G. & Schou, J. 2000. *apj* 537, 1026.

Emilio, M., Kuhn, J.R., Bush, R.I. & Scherrer, P. 2000. *apj* 543, 1007-1010.

Emilio, M., Couvidat, S., Bush, R. I. Kuhn, J. R. & Scholl, I. F. 2015. *apj* 798, 48 (8p).

Fazel Z., Rozelot, J.P., Lefebvre S., Ajabshirizadeh, A. & Pireaux, S. 2004. *Mem. Soc. Astro. It*. Vol. 75, 282-285.

Fazel, Z., Rozelot, J.P., Lefebvre, S., Ajabshirizadeh, A. & Pireaux, S. 2008. *newas* 13, 65-72.

Fiala, A.D., Dunham, D.W., Sofia, S., 1994. *solphys* 152, 97.

Fivian, M., Hudson, H. & Krucker, S. 2016. *Am. Astro. Soc.*, SPD meeting 47, id.12.04. Bibliographic Code:   2016SPD....47.1204F

Giannuzi, M.A. 1953. *Mem. Soc. Astro. It.*, 305-314.

Giannuzi, M.A. 1955. *Mem. Soc. Astro. It.*, 447-454.

Giménez de Castro, C. G., Varela Saraiva, A. C., Costa, J. E. R. & Selhorst, C. L. 2007. *aap* 476, 369.

Godier, S. & Rozelot, J.P. 2001. *solphy*s 199, 217–-229.

Goode, P.R. & Dziembowski, W.A., 2003. *Jour. of the Korean Astron. Soc.* 36, S75-S81.

Haberreiter, M., Schmutz, W. & Kosovichev, A.G. 2008. *apj* 675, L53–-L56.

Ragadeepika Pucha, Hiremath, K. M., Shashanka, R. & Gurumath, 2015. *JOA* "Development of a Code to Analyze the Solar White-light Images From the Kodaikanal Observatory: Detection of Sunspots, Computation of Heliographic Coordinates and Area".

Hudson, H. & Battaglia, M. 2014. http://sprg.ssl.berkeley.edu/~tohban/wiki/index.php/The_Solar_X-ray_Limb_II

Hudson, H. & Rozelot, J.P. 2010. http://sprg.ssl.berkeley.edu/~tohban/wiki/index.php/History_of_Solar_Oblateness

Iorio, L., Lichtenegger, H.I.M., Ruggiero, M.L. & Corda, C. 2011. *apss* 331, 351-–395.

Kilcik, A., Sigismondi, C., Rozelot J.P. & Guhl, K. 2009. *solphys* 257, 237–-250.

Kosovichev, A.G. & Rozelot, J.P. 2016. In ``*ISSI/VarSITI FORUM on Expected Evolution of Solar Activity in the Following Decades*''. Bern (CH) (1-3 March 2016), Georgieva, K. (ed). To be published.

Kosovichev, AG. & Zhao J. 2016. In ``*Cartography of the Sun and Stars*''. Lecture Notes in Physics, Vol. 914, Rozelot J.P., Neiner, C. (eds). Springer Verlag, Berlin, 25-41.

Kosovichev, A. 2011. In ``*The Pulsations of the Sun and the Stars*. Lecture Notes in Physics, Vol. 832. Rozelot, J.P. and Neiner, C. (eds). Springer-Verlag, Berlin.

Krivova, N. A., Solanki, S. K., Fligge, M., & Unruh, Y. C. 2003. *aap* 399, L1.

Kuhn, J.R., Bush, R.I., Emilio, M. & Scherrer, P.H. 2004. *apj* 613, 1241-1252.

Laclare, F., 1983. *aap* 125, 200-203.

Lamy, P., Prado, J.Y., Floyd, O., Rocher, P., Faury, G. & Koutchmy, S. 2015. *solphys* 290, 2617-2648.

Lefebvre, S., Kosovichev, A.G., Nghiem, P., Turck-Chièze, S. & Rozelot, J.P. 2006. In "*SOHO 18 Conference"*,Sheffield, U.K., August 7-11, 2006. ``Beyond the Spherical Sun: a new era of helio-and asteroseismology". ESA-SP, 624, CDROM, 9.1

Lefebvre, S., Kosovichev, A.G. & Rozelot, J.P.  2007. *apj* 658, L13.

Lefebvre, S., Nghiem, P.A.P. & Turck-Chièze, S. 2009. *apj* 690, 1272-1279.

Lefebvre, S., Bertello, L., Ulrich, R. K., Boyden, J. E. & Rozelot, J.P. 2006. *apj* 649, 444-451.

Lefebvre, S. & Kosovichev, A.G. 2005. *apj* 633, L149.

Lejeune, A. 1947, *Annales Société Scientifique de Bruxelles*, Série 1, Vol. LXI, 27, 27-47.

Livingston, W., Gray, D., Wallace L. & White, O.R., 2005. In ``*Large-scale Structures and their Role in Solar Activity*'' ASP Conference Series, Vol. 346, Proceedings of the Conference held 18-22 October, 2004 in Sunspot, New Mexico, USA, K. Sankarasubramanian, M. Penn & A. Pevtsov (eds), p. 353.

Meftah, M., Corbard, T., Irbah, A., et al. 2014. *aap* 569, A60.

Meftah, M., Hauchecorne, A. Irbah, A. et al. 2015. *apj* 808, 16p.

Meftah, M., Hauchecorne, A. & Irbah, A. 2016. *adv  58*, Issue 7, 1425–1440.

Morand, F., Delmas, C. Chauvineau, B. et al. 2010. *Comptes Rendus Acad. Sc. Phys.*, 11, 660.

Neckel, H. & Labs, D. 1987. *solphys* 110, 139-170.

Neckel, H. & Labs, D. 1994. *solphys* 153, 91-114.

Pap, J., Rozelot, J.P., Godier, S. & Varadi, F. 2001. *aap* 372, 1005-1018.

Pierce, A. K. & Slaughter, C. D. 1977-a. *solphys* 51, 25.

Pierce, A.K., Slaughter, C.D. & Weinberger, D., 1977-b. *solphys* 52, 179.

Pipin, V.V. & Kosovichev, A.G. 2011. *apj* 727, L45.

Reiter, J., Rhodes Jr., E. J., Kosovichev, A. G., Schou, J., Scherrer, P. H. & Larson, T. P. 2015. *apj* 803-92, 42 pp.

Rösch, J. & Yerle, R. 1983. *solphys* 82, 139-150.

Rozelot, J.P., Kosovichev, A.G. & Kilcik, A. 2015. *apj* 812, 4pp.

Rozelot, J.P., Pireaux, S., Kosovichev, A. & Damiani, C. 2009. In ``*Highlights of Astronomy*'', MEARIM Regional IAU meeting, Cairo (Egypt), Athem W. Alsabti, Ahmed Abdel Hady & Volker Bothmer (eds), University of Cairo Press. DOI: 10.1017/977403330200170.

Rozelot, J.P. & Damiani, C. 2012. *The Euro. Phys. Jour. H*. 37, 709-743.

Rozelot, J.P., Kosovichev, A.G. & Kilcik, A. 2016. In ``*Solar and Stellar Flares and Their Effects on Planets*'', Proceedings IAU Symposium No. 320, A.G. Kosovichev, S.L. Hawley & P. Heinzel (eds.), Cambridge University Press (London, UK), 9pp.

Scherrer, P.H., Bogart, R. S., Bush, R. I., Hoeksema, J. T., Kosovichev, A. G., Schou, J. et al. 1995. *solphys* 162, 129-188.

Schou, J., Kosovichev, A. G., Goode, P. R., & Dziembowski, W. A. 1997. *apj* 489 L197-L200.

Schou, J., Kosovichev, A.G., Goode, P.R. & Dziembowski, W.A. 1997. *apj* 489, 197.

Secchi, A. & Rosa, P. 1872. *Comptes Rendus Acad. Sc*. Vol. LXXIII and LXXIV.

Selhorst, C.L., Silva, A. V. R., & Costa, J. E. R. 2004. *aap* 420, 1117

Shapiro, A.E. 1975. *Journal of Historical Astronomy* 75-80.

Sigismondi, C., Ayiomamitis, A., Wang, X., Xie W., Carinci, M. & Mimmo, A. 2015. http://de.arxiv.org/ftp/arxiv/papers/1507/1507.03622.pdf

Sofia, S., Maier, E. & Twigg, L. 1991. *adv* Vol. 11, 4, 123-132.

Sofia, S., Girard, T. M., Sofia, U. J. et al. 2013. *mnras* 436, 2151.

Sukhodolov, T., Rozanov, E., Ball, W.T. and 15 co-authors 2016. *jgr Atmos*. 121, 6066-6084.

Ulrich, R.K. & Bertello, L. 1995. *Nature* 377(6546), 214-215.

Vaquero, J.M., Gallego, M.C., Ruiz-Lorenzo, J.J., López-Moratalla T., Carrasco, V.M.S., Aparicio, A.J.P., González-González, F.J. & Hernández-García, E. 2016. *solphys* 291, 1599-1612.

Wittmann, A. 1997. *aap* 61, 225-227.

Wittmann, A.D. & Bianda, M. 2000. In ``*The solar cycle and terrestrial climate, Solar and Space Weather Euroconference*'', Santa Cruz de Tenerife, Tenerife, Spain, *ESA SP*, A. Wilson, A. (ed). Vol. 463, 113-116, Noordwijk (NL).